%% file: ms.tex
\documentclass[12pt,preprint]{aastex}
\usepackage{natbib,graphicx,latexsym}
\newcommand{\lsim}{\
\raise-2.truept\hbox{\rlap{\hbox{$\sim$}}\raise5.truept\hbox{$<$}\ }}
\newcommand{\gsim}{\
\raise-2.truept\hbox{\rlap{\hbox{$\sim$}}\raise5.truept\hbox{$>$}\ }}

\newcommand{\vh}{$(V{-}H)$}
\newcommand{\vk}{$(V{-}K)$}
\newcommand{\ct}{$(C{-}T_1)$}
\renewcommand{\ur}{$(U{-}R)$}
\renewcommand{\bv}{$(B{-}V)$}
\newcommand{\bi}{$(B{-}I)$}

\newcommand{\vi}{$(V{-}I)$}
\newcommand{\ik}{$(I{-}K)$}
\newcommand{\ih}{$(I{-}H)$}
\newcommand{\jk}{$(J{-}K)$}

\begin{document}

\shorttitle{Metallicity-Color Relations in GC Systems}
\shortauthors{Cantiello \& Blakeslee}

\title{On the Metallicity-Color Relations and Bimodal Color
Distributions in Extragalactic Globular Cluster Systems}

\author{Michele Cantiello\altaffilmark{1,2}}
\and
\author{John P. Blakeslee\altaffilmark{1}}

\altaffiltext{1}{Department of Physics and Astronomy, Washington State University,
Pullman, WA 99164.}
\altaffiltext{2}{INAF--Osservatorio Astronomico di
Teramo, Via M. Maggini, I-64100 Teramo, Italy}

\begin{abstract}
We perform a series of numerical experiments to study how the nonlinear
metallicity--color relations predicted by different stellar population
models affect the color distributions observed in extragalactic globular
cluster systems.  
We present simulations in the $UBVRIJHK$ bandpasses
based on five different sets of simple stellar population (SSP) models.
The presence of photometric scatter in the colors is included as well.
We find that unimodal metallicity distributions frequently ``project''
into bimodal color distributions.  The likelihood of this effect depends
on both the mean and dispersion of the metallicity distribution,
as well as of course on the SSP model used for the transformation.
Adopting the Teramo-SPoT SSP models for reference, we find that
optical--to--near-IR colors should be favored with respect to other
colors to avoid the bias effect in globular cluster color
distributions discussed by \citet{yoon06}.  In particular, colors such
as \vh\ or \vk\ are more robust against nonlinearity of the
metallicity--color relation, and an observed bimodal distribution
in such colors is more likely to indicate a true underlying bimodal
metallicity distribution.  Similar conclusions come from the
simulations based on different SSP models, although we also identify
exceptions to this result.
\end{abstract}


\section{Introduction}

The globular cluster (GC) system of a galaxy provides a
unique view into the formation history of the galaxy. 
Apart from some rare exceptions, GCs are known to represent a 
relatively simple class of objects, with homogeneous ages
and chemical compositions for the stars composing each GC. Thus, GCs
are at the same time reasonably simple objects, and good tracers
of the early star formation histories of the host galaxy.
For these reasons, in recent years a great deal of effort has been made
to study the observational properties of extragalactic GC systems. As
a consequence, many GC features have been discovered, providing
valuable constraints on the evolutionary paths of galaxies.

One commonly observed property is that the GC populations in galaxies
tend to be bimodal in their color distributions.  A combination of
photometric and spectroscopic observations indicates that GC systems
are fairly homogeneous in terms of age, so differences in color mainly
reflect metallicity differences.  Thus, the bimodal color
distributions have usually been interpreted as bimodal metallicity
distributions; see the reviews by \citet{west04}, \citet{brodie06},
and references therein, for further details.
However, more recently some authors
\citep[e.g.][]{richtler06,yoon06} have shown that a bimodal color
distribution can be enhanced, or even originated, by the effect of 
nonlinear metallicity-color (MC hereafter) relations.

For instance, \citet{richtler06} has shown that the observed MC
relation for the Washington system \ct\ color, coupled with a
Gaussian scatter of 0.08 mag around the mean relation, can transform a
nearly flat GC metallicity distribution into a bimodal \ct\
distribution. \citet[][YYL06 hereafter]{yoon06}, instead, exploited
the fact that, when the Horizontal-Branch (HB) morphology is
realistically modeled in stellar population simulations, the color
indices sensitive to stars in this evolutionary stage follow ``wavy''
MC relations. Such a nonlinear feature has in fact been observed for the
metallicity versus $(g-z)$ color of GCs in the Milky Way and Virgo
ellipticals \citep{peng06}. This feature causes evenly-spaced
metallicity bins to be ``projected'' into larger/narrower color bins,
depending on the location on the MC relation. YYL06 consequently conclude
that it is not necessary to invoke a
bimodal metallicity distribution to have a bimodal color
distribution.  We will refer to this effect
as {\it metallicity  projection bias}.

In this paper we study how nonlinear MC relations can affect the
color distributions observed in different passbands. Our aim is  ($i$)\,to test how
various colors ``suffer'' from the nonlinear effects described above,
and, consequently, ($ii$)\,to suggest the optimal color(s) for revealing
the presence of real bimodal GC metallicity distributions.

We do this by first carrying out multiple sets of simulations based on the
Simple Stellar Population (SSP) models developed by the Teramo-SPoT group \citep[][SPoT
hereafter\footnote{The SPoT models are available at the Teramo-SPoT website:
www.oa-teramo.inaf.it/SPoT}]{raimondo05}. We then perform the same tests using
simulations based on four other sets of SSP models and compare the results.
Finally, we summarize the most robust conclusions on GC
colors and their underlying metallicity distributions from this work.

\section{Models simulations}
For this study
we adopt the SPoT SSP models as our reference models for two reasons.
First, these models have proven to match fairly well
the observed integrated photometric properties
of galaxies, i.e. colors, surface brightness fluctuation magnitudes,
etc., in different passbands for a large sample of objects with very
different physical properties
\citep{cantiello05,raimondo05,cantiello07}.
Second, the SPoT models also provide a good match to the observed
color--magnitude diagrams (CMDs) for star clusters with a wide range of
ages and chemical compositions \citep{brocato00,raimondo05}. In
particular, these models are optimized to simulate the HB spread
observed in Galactic GCs.

The detailed numerical synthesis of the CMD features is a key point
for the aims of the present study since, as shown by YYL06, the wavy
feature that can produce a ``projected'' color bimodality is due to
a realistic treatment of the HB morphology. It is worth noting here
that this feature is {\it not} a peculiarity of the YYL06 models; in
fact it was already presented by \citet[see their Figures 2 and
4]{lee02} and, as we will show, it is present also in the SPoT
models.  Not surprisingly, this feature is not noticeable in those
models where the HB morphology is not properly simulated to match the
observed Galactic GC properties, as for example in the \citet{bc03}
models, which adopt a fixed red HB morphology.

In fact, our reference SPoT models attempt to simulate in a realistic
way all features of the observed CMD, that is all the stellar
evolutionary stages including the fast and bright phases of the Giant
Branch stage.  These models are computed according to the following
prescriptions: \citet{scalo98} Initial Mass Function; solar
scaled stellar evolution tracks from \citet{pietrinferni04}; HB
morphology reproduced taking into account the effects due to age,
metallicity, and the stellar mass spread due to the stochasticity of
the mass-loss along the RGB. The RGB mass--loss rate is evaluated
according to Reimers' law \citep{reimers75}, with efficiency
$\eta_{RGB}=0.4$. Thermal pulses are simulated using the analytic
formulations by \citet{wagenhuber98}. Finally, the atmosphere models
are from \citet{westera02}. See \citet{raimondo05} for further
details.
Throughout this paper, we will consider the $t=13$ Gyr age models for
reference, if not stated otherwise.

\begin{figure}
\epsscale{1.}  \plotone{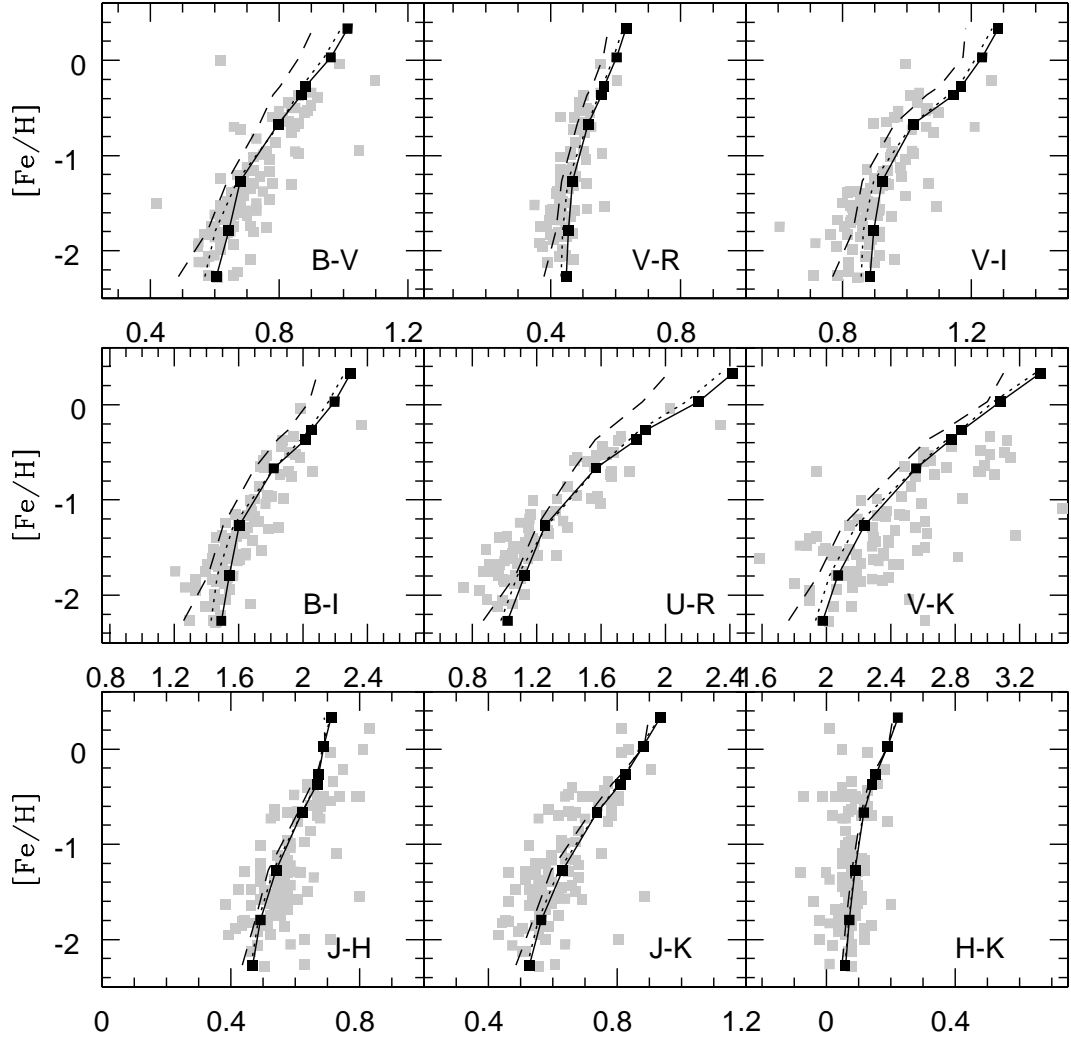}
\caption{The Teramo-SPoT models compared to observational data. The
models refer to three different ages: 5 Gyr (dashed lines), 11 Gyr
(dotted lines) and 13 Gyr (solid lines, reference models).  The gray
dots mark Galactic GC data.
\label{cmspot}}
\end{figure}

Figure~\ref{cmspot} shows the MC relations from the SPoT models for
several different colors.  Data for the Galactic GCs are also
shown. The optical colors and the [Fe/H] values for
the Galactic GCs are taken from the \citet{harris96} updated online
catalog\footnote{http://www.physics.mcmaster.ca/$\sim$harris/mwgc.dat},
while the near-IR photometric data are from \citet{brocato90} and
\citet{cohen07}.  As seen in Fig.~\ref{cmspot}, the models provide 
a good match to the integrated properties of the Galactic GC system. 
The ``wavy'' behavior of the MC relations for the \vi,
\bi, \ur, and \bv\ colors is clearly evident.
Furthermore, it is worth emphasizing the general nonlinearity
of the MC relations for all the colors shown in the figure.

\subsection{GC Simulations: SPoT models}

We have developed a procedure to simulate a GC population with an
arbitrary metallicity distribution and number of objects. Throughout
this paper, however, we will consider the case of Gaussian metallicity
distributions, and we simulate GC populations composed of 1000 objects.
Armed with the MC relations of our reference models, the metallicity
distribution of the GC system is randomly populated and projected into
a color distribution. Finally, we use the KMM code
\citep{mclachlan88,ashman94} to test whether the GC color distribution is
best fit by a single or double Gaussian function.

\begin{figure}
\epsscale{1.1}
\plottwo{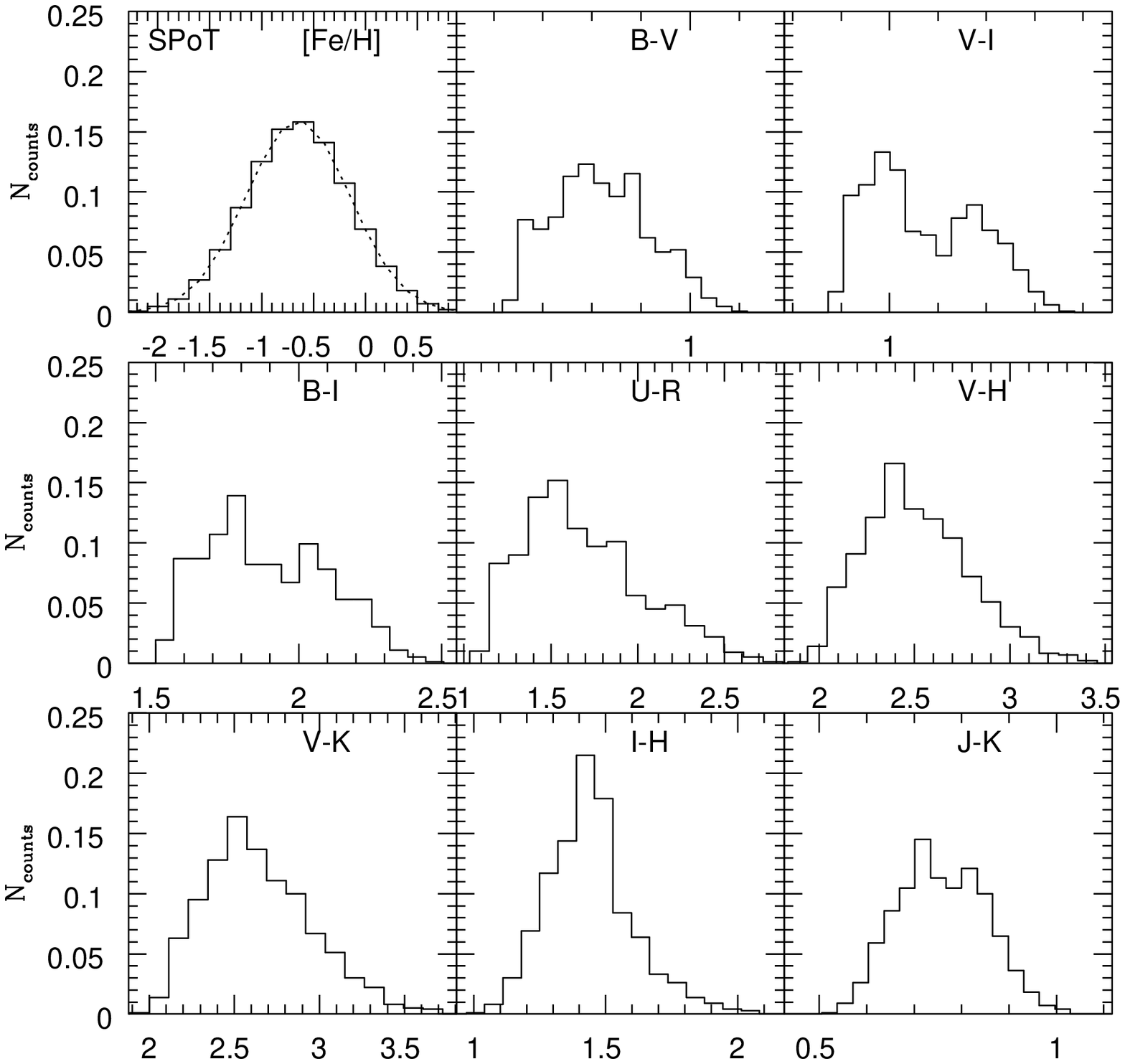}{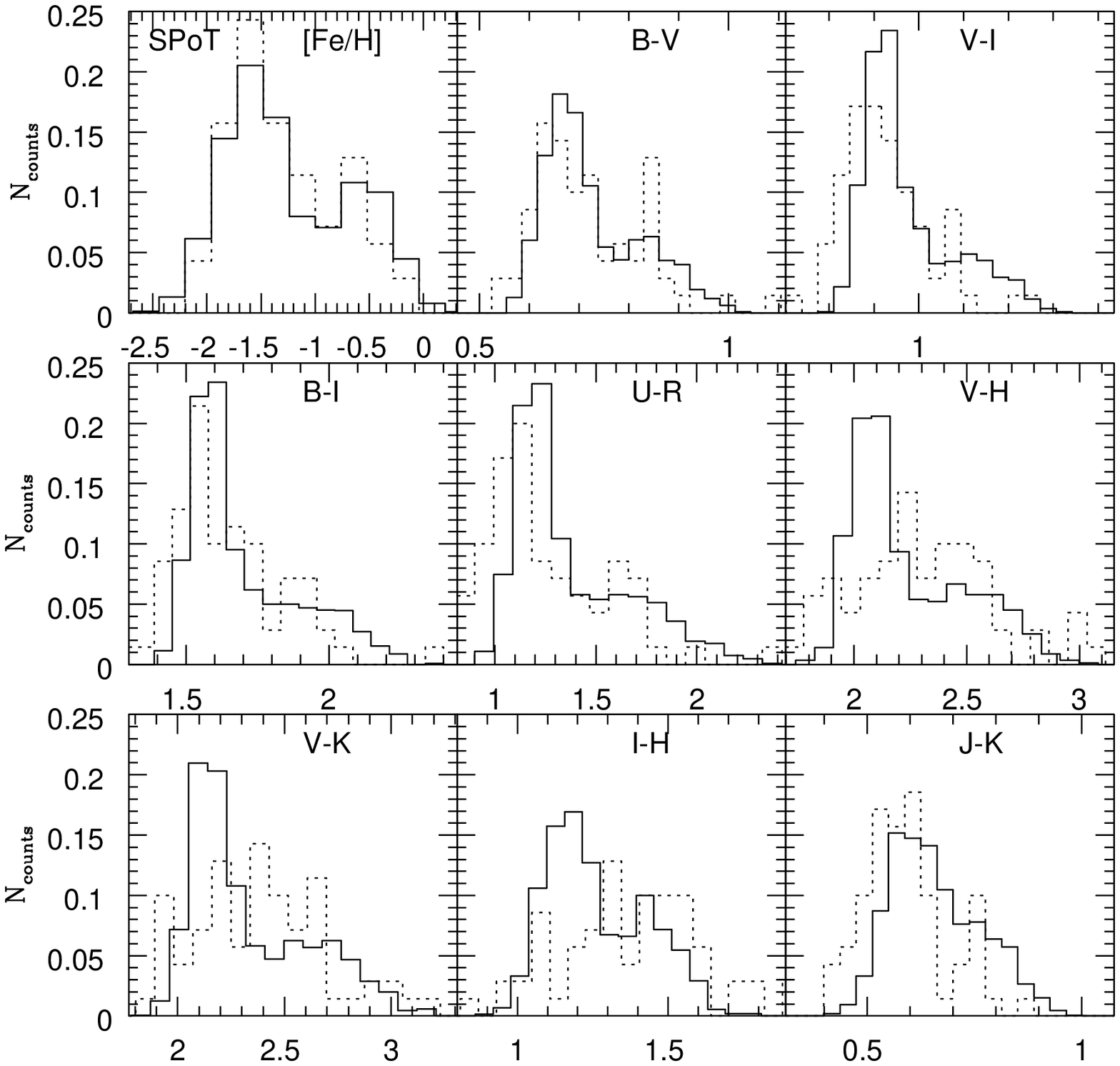}
\caption{\textit{Left panel:} The color histograms obtained from a
unimodal (Gaussian) metallicity distribution with mean
[Fe/H]$\,=-0.65$, $\sigma_{\rm [Fe/H]}=0.5$ (shown in the upper left
panel). We find that all the optical and the \jk\ color distributions
are bimodal based on the KMM algorithm, while for the other colors a
double Gaussian distribution does not significantly improve the fit to
the data.  \textit{Right panel:} Simulation of a bimodal metallicity
distribution (solid lines), chosen to match the Galactic GC
distribution. The observed color histograms for the Galactic GCs
are also shown with dotted lines (the observed metallicity
distribution is shown in the upper left panel with dotted histogram).
The parameters used for the simulated distribution are
[Fe/H]$_{low}=-1.59$, $\sigma_{[Fe/H],low}=0.30$ dex, and
[Fe/H]$_{high}=-0.55$, $\sigma_{[Fe/H],high}=0.27$ dex, with a
photometric uncertainty of 10\%, and the N$_{low}$ to N$_{high}$ ratio
is 2.
\label{bimodal}}
\end{figure}

In Figure \ref{bimodal} (left panel), we show the results of one of
these simulations. Specifically, in this case we have simulated a
metallicity distribution similar to the one adopted by YYL06, that is
a Gaussian with peak at [Fe/H]$\,=-0.65$ dex and dispersion
$\sigma_{[Fe/H]}=0.5$ dex. It is clearly recognized from the \vi\
panel that the projected color distribution is bimodal.  By running
the KMM code, we find that, for this specific simulation, all
the optical colors\footnote{We consider the \ur\ as it is the
nearest color the Washington system \ct\ color, not provided
with the SPoT models. The \ct\ index is interesting because it is
known to be one for which the GCs distribution is bimodal in all of the
limited number of observed galaxies \citep{richtler03}.}  and the \jk\ color
distributions are significantly bimodal, while the optical to
near-infrared colors, including \ih, have unimodal distributions.

As a check to these simulations, we have also made some numerical
experiments adopting a bimodal metallicity distribution. In
particular, Figure \ref{bimodal} (right panel) shows a simulation
carried out adopting the bimodal metallicity distribution of the
Galactic GC system, obtained using the prescriptions of
\citet{cote99}. In the Figure also the observed metallicity and color
distributions of Galactic GCs are shown. The histograms of the
simulated GC population are shown with solid lines in the panels,
while the histograms for the actual observed Galactic GCs are shown
with dotted lines. To simulate observational scatter of the data, we
have included a 10\% Gaussian scatter in the colors. As can be seen
from this comparison, there is generally a good match between the
simulated and actual color histograms of the Galactic GC system.

Since our goal here is to identify the colors least affected by the
projection bias, regardless of the underlying metallicity
distribution, we have performed various tests assuming unimodal
Gaussian distributions with peaks at [Fe/H]$\,=-1.65, -1.15, -0.65,
-0.15$ dex and three values for the dispersion: $\sigma_{[Fe/H]}=0.25,
0.5, 0.75$ dex.  For each of the twelve ([Fe/H], $\sigma_{[Fe/H]}$)
pairs, we have simulated a GC system with a unimodal metallicity
distribution and evaluated the colors of each GC according to the
adopted MC relations.  Afterwards, by using the KMM code, we estimate
the likelihood, $P(bimodal)$, that the color distribution is better
represented by two Gaussians than a single Gaussian, for various color
choices. Values of $P(bimodal){\,\approx\,}1$ mean that the color
distribution is likely bimodal; conversely, color distributions with
$P(bimodal){\,\approx\,}0$ are likely unimodal.  We have run the
simulations both with and without including a 10\% Gaussian scatter in
the simulated colors.

\begin{figure}
\epsscale{1.}
\plotone{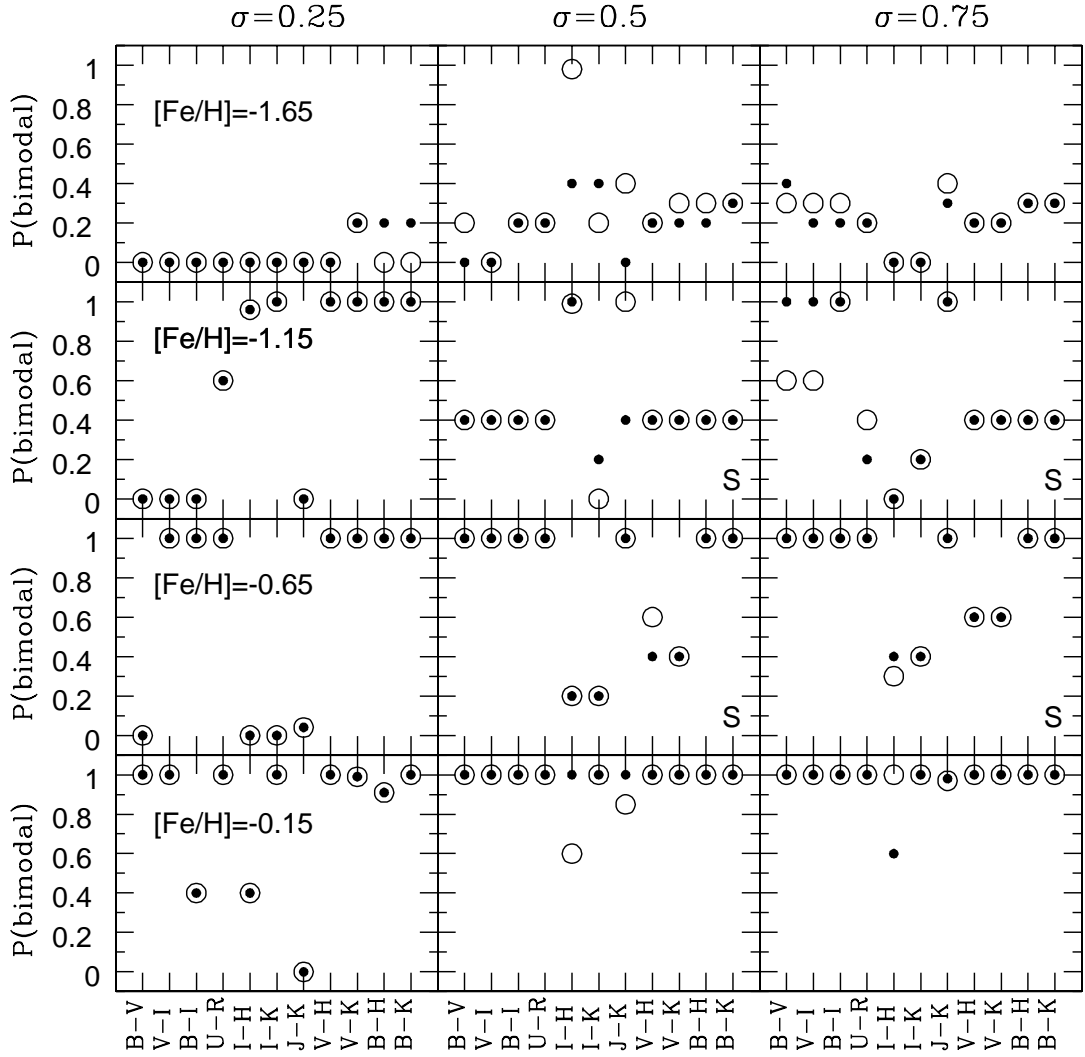}
\caption{The probability $P(bimodal)$ of having a bimodal color distribution
starting from a unimodal Gaussian metallicity distribution is
shown for different color indices and various metallicity 
distributions.   Left/middle/right panels refer to
simulations with $\sigma=0.25,0.5,0.75$ dex, respectively (see upper
labels). Different rows refer to different mean metallicities, as
labeled. High\,(low) values of $P(bimodal)$ mean that the color distribution
is significantly bimodal\,(unimodal). Filled dots mark numerical experiments without any
color scatter, and open circles show simulations including a 10\% color
scatter. Although all the simulated metallicity distributions are unimodal,
about 45\% of these color distributions are found to be bimodal.
The four panels with the ``\textsf{S}''  label refer to the simulations
that best match with observed GC color ranges.
\label{plotspot}}
\end{figure}

Table \ref{tab_spot} gives the results of these simulations. 
For each color index, the table lists the locations of the best-fitting 
blue and red peaks and the value of $P(bimodal)$
for each choice of mean metallicity and dispersion. The results are 
also shown graphically in the Figure \ref{plotspot}, where solid dots
mark the results for simulations without any color errors, and
open circles mark results obtained with the random 10\% color scatter.
The different rows and columns refer to
different mean [Fe/H] and $\sigma$ values, respectively, as labeled.

Two considerations emerge from inspection of Figure \ref{plotspot}.
First, according to the SPoT models, the projection effect that
causes a unimodal metallicity distribution to be observed as a bimodal
color distribution is not a unique characteristic of the HB-sensitive
colors.  It is, instead, present for most of the analyzed colors. For
example, in the case of \jk, almost half of the numerical
experiments carried out give bimodal color distributions
[$P(bimodal)\sim 1.0$].  Thus the nonlinearity of the MC relation
is not specific to just one or a few colors, 
such as the $(g-z)$ and \vi\ colors discussed by YYL06. 
Although, as shown in Figure \ref{cmspot}, different colors are
affected differently by nonlinearity in the MC relation.

The second consideration that emerges from these simulations regards how
the presence of color scatter (i.e. the photometric uncertainty) can
affect the probability of obtaining a bimodal color distribution.  
The addition of color scatter can of course decrease the probability 
of bimodality by smoothing 
out the separation between the peaks.  More surprisingly, it can also
make bimodality appear more probable by removing sharp features from
the color distributions, and thus significantly improving the 
goodness-of-fit of the double Gaussian model used by KMM.

It is interesting to note that the two extreme metallicities,
[Fe/H]$\,=-1.65$ and $-$0.15 dex, result in strictly-unimodal, and
generally-bimodal, color distributions, respectively \citep[see
also][their Fig. 3]{yoon06}. Moreover, simulations with larger
$\sigma$ values are in almost all cases more bimodal.  Thus, for the
combination ([Fe/H]$\,=-$0.15, $\sigma{\,=\,}0.75$), the color
distributions are significantly bimodal for all simulated colors, but
again this is an extreme case.
The color distributions (as indicated by the peaks in Table \ref{tab_spot})
obtained from the most extreme simulations are not typical of those
normally observed for extragalactic GC systems.

In order to refine our study, we now focus on those simulations best 
matching real GC systems.   We have compared the color peaks from Table
\ref{tab_spot} with observed color peaks from literature.  In
particular, we have selected as ``realistic'' the simulations with:
\begin{itemize}
\item $(V-I)_{0,blue} \sim 0.95$ and $(V-I)_{0,red}\sim 1.15$, based
on the \citet{brodie06} compilation 
for bright (mainly E and S0) galaxies with $M_B \leq -18.5$~mag.

\item $(B-I)_{0,blue} \sim 1.94$ and $(B-I)_{0,red} \sim 2.06$,
derived from the \citet{harris06} sample of bright galaxies.

\item $(I-H)_{0,blue} \sim 1.3$ and $(I-H)_{0,red} \sim 1.7$,
from \citet{kundu07}, based on M\,87.

\end{itemize}

In order to avoid any bias towards bimodal distributions, we have also
considered as realistic those unimodal color distributions whose peak
is equal to the averaged blue and red peak colors reported above. By
matching these criteria with the simulations, we have found that only
the subset of simulations with [Fe/H]$\,=-$1.15, $-$0.65 
and $\sigma_{[Fe/H]}=0.5, 0.75$
provide realistic ranges for the color peaks.  For example, the peaks
for the \vi, \bi, and \ih\ colors for the case of ([Fe/H]$\,=-$1.15,
$\sigma_{[Fe/H]}=0.75$) are all in good agreement with the observational
values listed above, even though most of these colors are found to
have unimodal color distributions for this particular simulation.

By inspecting only the panels for the 
\textit{\sffamily s}elected ``realistic'' simulations in
Figure~\ref{plotspot} (the panels labeled with an ``\textsf{S}''),
one can see that the colors \bv, \vi, and \bi\ have, on average,
an increased probability of being projected to a bimodal distribution,
while colors such as \ik, \vh and \vk\ have lower probabilities.
Thus, if one wants to minimize the bias from the MC
projection effect in real observations, i.e. if the
contribution to bimodality due to a nonlinear MC relation
{is} to be neglected, then the \ik, \vh, and \vk\
colors are to be preferred.

Finally, we must emphasize that the above conclusions do not change if
we adopt different ages. In fact, although there is some shift in
color, the MC {\it profiles} are not strongly affected even when the
$t=5$ Gyr models are considered (Figure \ref{cmspot}).  Since the
projection effect is due to the shape of the MC relation (that is, to
the changing derivative of the relation), and not to the absolute
color values, this explains why the outcome of the simulations does
not change significantly with age.  In more detail, for the
\citet{raimondo05} models  at an age of $t=5$ Gyr, the ``wavy'' MC
relation is mostly related to the appearance of the HB at
metallicities [Fe/H]$\lsim -0.4$ dex, while no HB is present at higher
[Fe/H].  Finally, it is worth noting that the above results do not
change substantially if the numerical experiments are carried out
using a different number of simulated GCs\footnote{We have found that
the locations of the color peaks change on average $\lsim$0.05 mag,
and $P(bimodal)$ by less than 25\% if $>$50 up to $\sim$2000 GC are
considered in the simulations. Numerical experiments with less than 50
sources can significantly deviate from the results given in Table
\ref{tab_spot}. Thus, our results should be compared with observations
that include color data for more than 50 GCs.}.

\subsection{Other SSP models}
The results presented in the previous Section are based on a
particular choice of the MC relations derived from the SPoT SSP
models. In order to verify the robustness of those results, in this
section we perform the same analysis discussed above, but with the MC
relations derived from four other sets of SSP models. We consider the
\citet{maraston05}, \citet{anders03}, \citet{bc03} and \citet{lee02}
models (hereafter Mar05, And03, BC03, and Lee02, respectively).

\begin{figure}
\epsscale{1.}
\plottwo{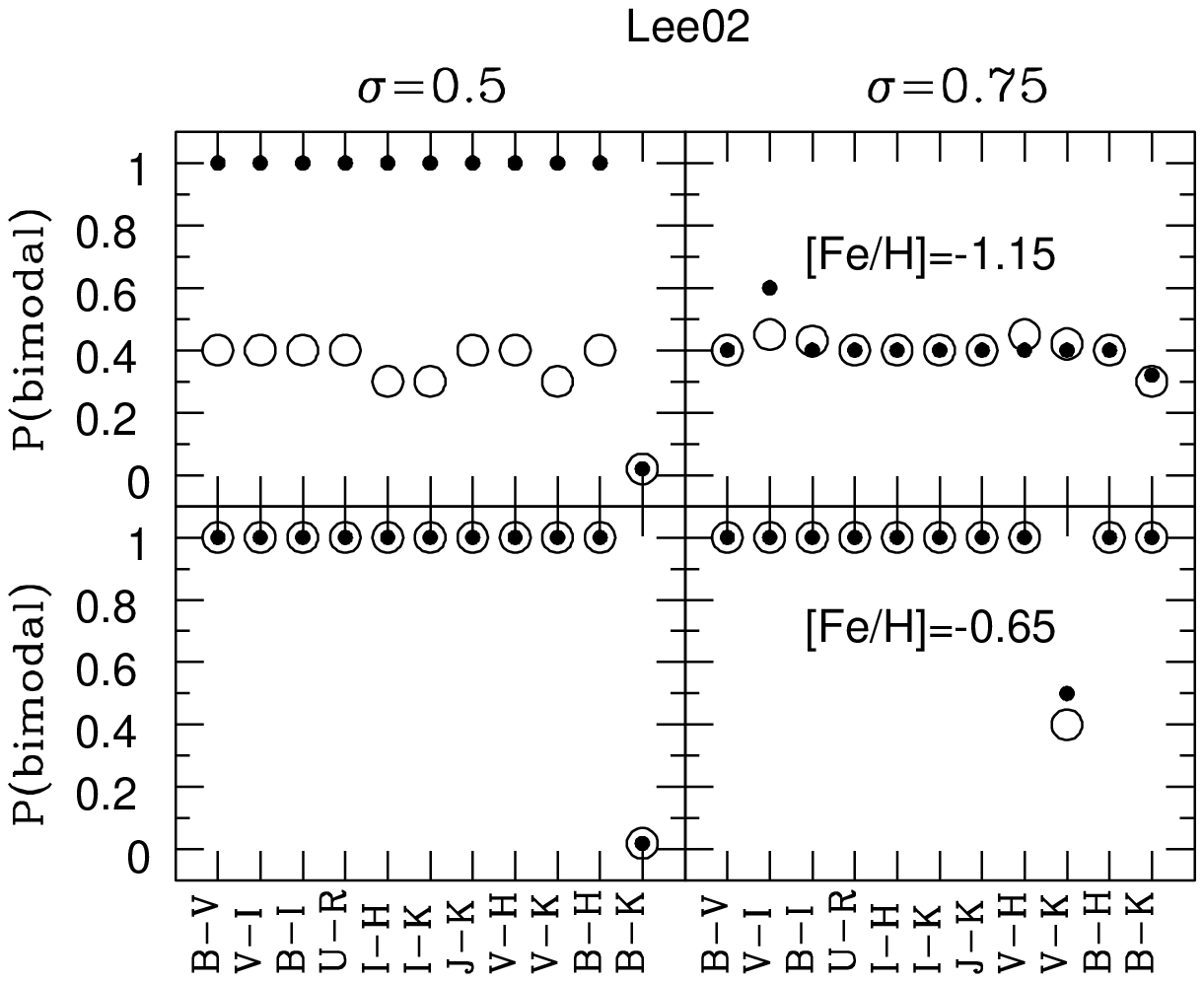}{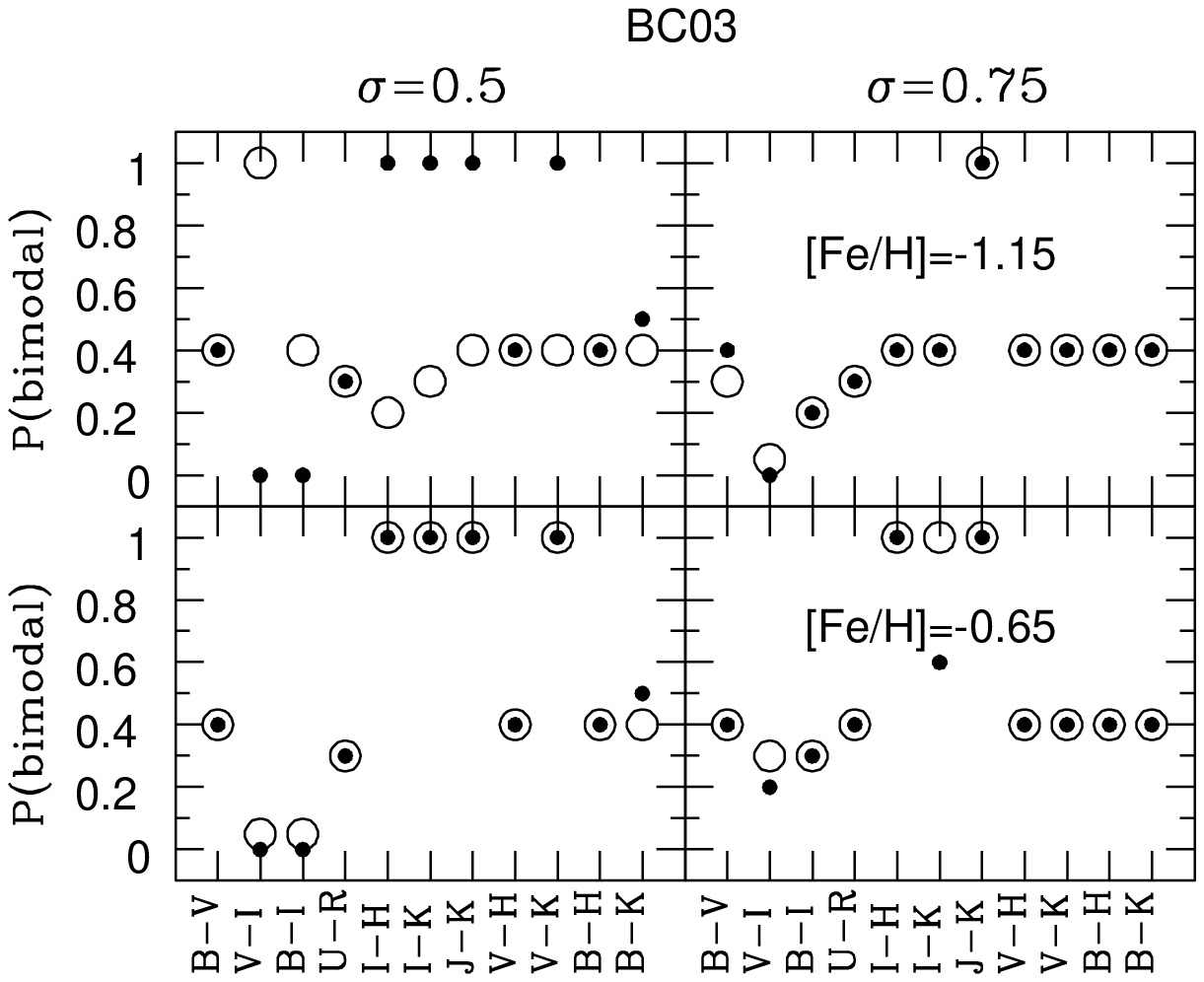}
\plottwo{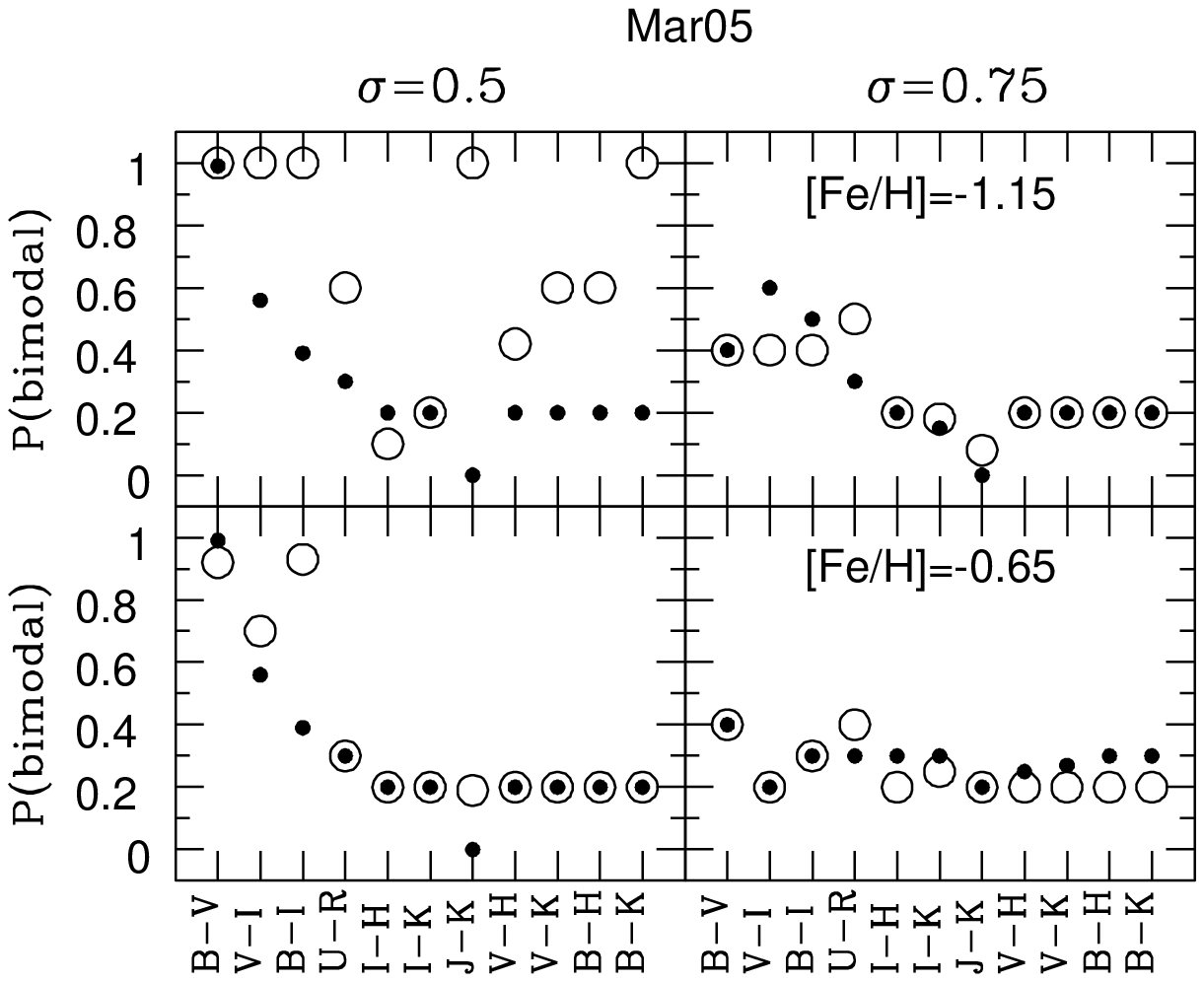}{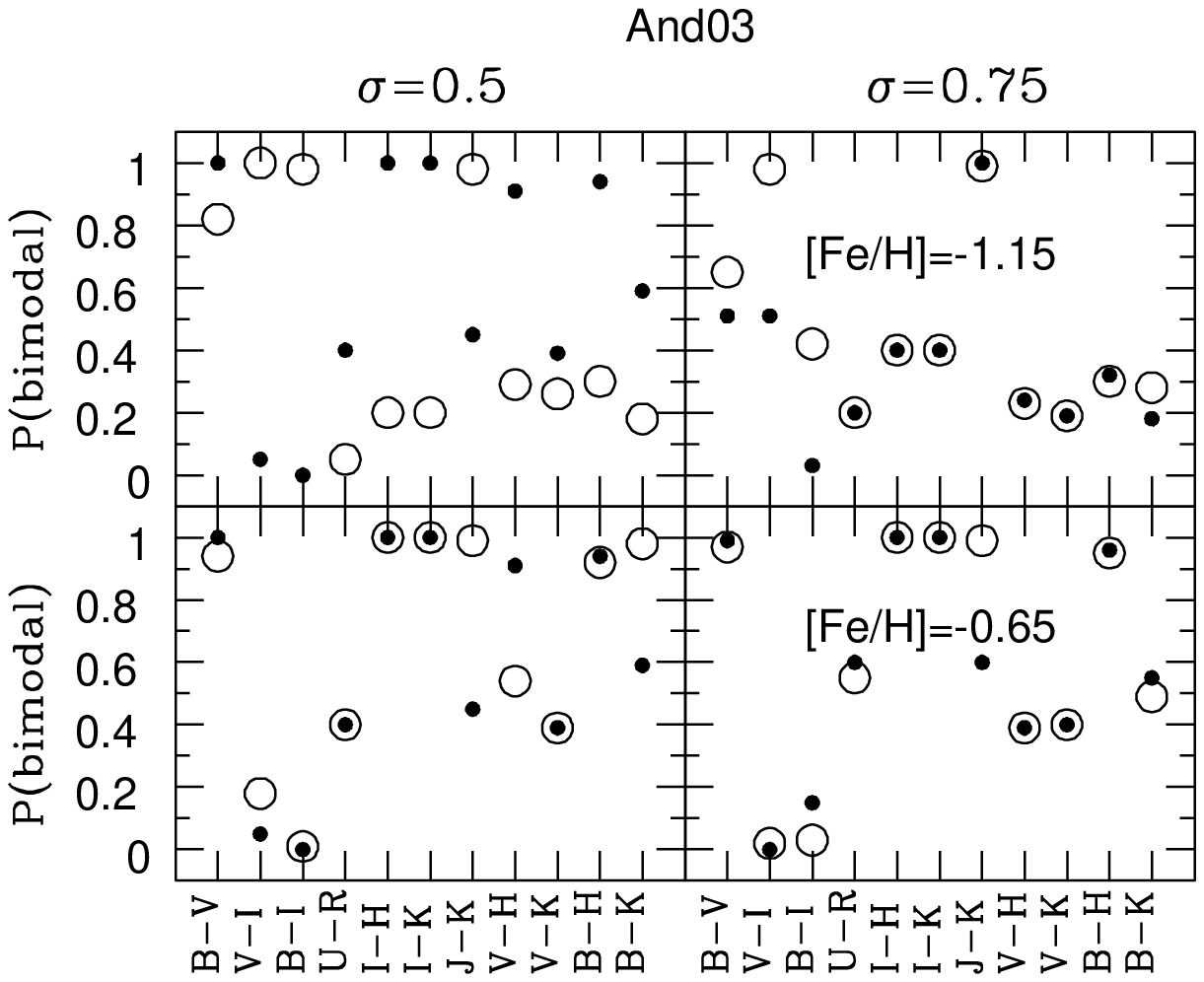}
\caption{Same as Figure \ref{plotspot}, but for MC relations taken from
other sets of SSP models (labeled at the top of each set of panels). 
Only the results for the simulations matching with observed GC colors are shown.
\label{plotabsel}}
\end{figure}

We emphasize that, with the lone exception of the Lee02 models, the
quoted models are computed with the primary aim of deriving the
integrated photometric properties of stellar systems. This means that,
in contrast to the SPoT models, they are not constrained to match 
as well with the specific features of observed CMDs. As a
consequence, the detailed shape of the MC relation may not take
into account the effect of stars in a particular evolutionary
phase, which is a key point for a detailed modeling of the MC
relations. Keeping in mind this warning, we perform for these models
the same analysis discussed above for the SPoT models.
For these simulations we again adopt a model age of 13 Gyr, except for the
Lee02 models, which do not include this age, so we use their 14~Gyr models.
The results of the simulations are presented in Figure \ref{plotabsel}, where we
show only the results for the selected ``realistic'' simulations, although this
choice does not substantially affect the our conclusions.

Inspecting the panels of Figure \ref{plotabsel}, we find some
differences with the results based on the SPoT models. For example, 
the values of $P(bimodal)$ are low for the BC03 \bv, \vi, and \bi\ colors
distributions.
This result for the BC03 models is not surprising, due to their lack of
detailed HB morphology modeling, which is the main cause of the wavy
MC relations for these colors.   In contrast, the \jk\ colors
from the BC03 models are almost always bimodal, a result of some
non-linearity in their MC relation unrelated to HB morphology.
On the other hand, the Lee02 models, where nonlinear effects in the
MC relation are stronger with respect to other models, generally
predict higher $P(bimodal)$ values.\footnote{We chose the 14 Gyr
Lee02 models specifically because the ``wavy'' feature is more pronounced;
this allows us to highlight better the influence of such features 
on the color distributions. The MC relations for the Lee02 preferred
12~Gyr models give results more similar to the SPoT ones.}.  

By making a cross-check of the results based on these sets of SSP
models with the ones based on the SPoT models, we find that no one
color is completely unaffected by MC projection bias. 
However, in almost all cases the \vh\ and \vk\ colors are predicted to be
less affected by this bias. Thus, these mixed optical--IR colors should be preferred
for GC studies, since, in normal galaxies, a bimodal distribution in these
colors is more likely linked to an underlying bimodal metallicity distribution.

\section{Conclusions}

In this work we have performed a series of numerical experiments to
simulate the properties of GC populations observed in different
photometric colors.  Our aim was to study how the nonlinear behavior of
the MC relations affect a unimodal (Gaussian) metallicity distribution
when it is projected into various optical and near-IR color distributions.
By using the MC relations from the SPoT models, we have found that
{\it a unimodal metallicity distribution can be projected into a
bimodal color distribution in almost any of the colors considered here,
depending on the properties of the metallicity distribution, on the
particular color index, and on the photometric uncertainty of the sample.}

This result is due to the fact that all the MC relations are by and large
nonlinear.  To reduce the possibility of this bias in real data, and thus
help ensure that an observed bimodal color distribution is due
to a bimodal metallicity, one should choose a color whose MC relation
is nearly linear. Since, for the grid of colors that we have
considered here, there is no such ``unbiased'' color, the best colors
to use are those that are most robust against this effect.
Using the SPoT models, we have concluded that optical--to--near-IR
colors are the best choices to disclose real bimodal
metallicity distributions.  

In order to assess model systematics and make more firm conclusions,
we have also investigated several other sets of stellar population models.
As a general result, the differences existing
between model predictions do not allow us to pick any color index as
safely unaffected by the metallicity projection bias.  However, all
models considered here, including the SPoT ones, predict that the bias
effect is reduced for \vk, \vh, and similar colors.
One other result of these simulations is that photometric uncertainties
can affect, in surprising ways, the probability of obtaining a bimodal
color distribution from the KMM algorithm.  Thus, decreasing the statistical
errors in real color data can help to avoid false detection of
significant bimodality.

Further information on metallicity bimodality can of course come from the
analysis of spectroscopic data for a significant number of GCs in galaxies
with observed color bimodality. However, such observations are time consuming,
and only feasible for relatively nearby objects.
In addition, certain spectroscopic indices may themselves be affected by
similar nonlinear relations with metallicity.

In conclusion, we confirm \vh\ and \vk\ as good colors to reveal (nearly)
unbiased bimodal metallicity distributions in extragalactic GC systems. 
Future data on large GC samples in individual galaxies, including
optical and near-IR photometry, as well as spectroscopy, coupled with
further advances in stellar population modeling, should finally resolve
this issue.  Until that time, the interpretation of bimodal color
distributions will remain, at least in part, ambiguous.

\acknowledgments

We thank the anonymous referee for helping us to improve this paper
with useful suggestions. We would like to thank Eric Peng, Pat C{\^
o}t{\' e}, and Gabriella Raimondo for useful comments.  This research
was supported by the NASA grant AR-10642, and the paper was completed
under the sponsorship of a INAF-OA Teramo grant.

\clearpage

\clearpage

\input{tab1}


\bibliographystyle{apj}
\bibliography{cantiello_jun07}

\end{document}

%% file: tab1.tex
\begin{deluxetable}{l|c|c|c|c|c|c|c|c|c|c|c}
\rotate
\tabletypesize{\tiny}
\tablewidth{0pt}
\tablecaption{Simulations of GC population}
\tablehead{\colhead{[Fe/H], $\sigma$} & \colhead{$(B-V)_0$\tablenotemark{a}} &  \colhead{$(V-I)_0$} & \colhead{$(B-I)_0$} & \colhead{$(I-H)_0$} & \colhead{$(I-K)_0$} & \colhead{$(J-K)_0$} &  \colhead{$(V-H)_0$} &  \colhead{$(V-K)_0$} & \colhead{$(U-R)_0$} &
\colhead{$(B-H)_0$} &  \colhead{$(B-K)_0$} \\
\colhead{(dex)} & \colhead{blue red P$_b$} & \colhead{blue red P$_b$} & 
\colhead{blue red P$_b$} & \colhead{blue red P$_b$} & \colhead{blue red P$_b$} & 
\colhead{blue red P$_b$} & \colhead{blue red P$_b$} & \colhead{blue red P$_b$} &
\colhead{blue red P$_b$} & \colhead{blue red P$_b$} & \colhead{blue red P$_b$}}
\startdata
 & \multicolumn{11}{c}{GC simulations without color scatter} \\
\tableline
$-$1.65 0.25 &     0.66 0.66 0.0 &     0.91 0.91 0.0 &     1.57 1.57 0.0 &     1.15 1.16 0.0 &     1.22 1.24 0.0 &     0.59 0.59 0.0 &     2.05 2.07 0.0 &     2.13 2.31 0.2 &     1.17 1.19 0.0 &     2.71 2.94 0.2 &   2.79 3.02 0.2 \\
$-$1.15 0.25 &     0.72 0.72 0.0 &     0.95 0.96 0.0 &     1.65 1.79 0.2 &     1.25 1.35 1.0 &     1.34 1.46 1.0 &     0.66 0.67 0.0 &     2.19 2.36 1.0 &     2.29 2.47 1.0 &     1.32 1.51 0.4 &     2.90 3.13 1.0 &   3.00 3.24 1.0 \\
$-$0.65 0.25 &     0.81 0.82 0.0 &     1.02 1.14 1.0 &     1.80 2.00 1.0 &     1.42 1.44 0.0 &     1.55 1.56 0.0 &     0.75 0.76 0.0 &     2.43 2.63 1.0 &     2.56 2.79 1.0 &     1.58 1.85 1.0 &     3.22 3.49 1.0 &   3.35 3.65 1.0 \\
$-$0.15 0.25 &     0.92 0.93 0.0 &     1.08 1.21 0.2 &     1.95 2.14 0.2 &     1.56 1.74 0.4 &     1.72 1.93 1.0 &     0.85 0.86 0.0 &     2.74 2.94 0.6 &     2.90 3.12 1.0 &     1.88 2.21 1.0 &     3.62 3.87 0.9 &   3.77 4.07 1.0 \\
\hline
$-$1.65 0.5  &     0.67 0.67 0.0 &     0.91 0.92 0.0 &     1.56 1.80 0.2 &     1.12 1.32 0.4 &     1.21 1.43 0.4 &     0.58 0.61 0.0 &     2.04 2.37 0.2 &     2.12 2.49 0.2 &     1.16 1.53 0.2 &     2.69 3.15 0.2 &   2.77 3.26 0.3 \\
$-$1.15 0.5  &     0.70 0.84 0.4 &     0.95 1.14 0.4 &     1.65 2.00 0.4 &     1.23 1.42 1.0 &     1.37 1.63 0.2 &     0.64 0.78 0.4 &     2.20 2.60 0.4 &     2.30 2.76 0.4 &     1.33 1.85 0.4 &     2.90 3.44 0.4 &   3.00 3.60 0.4 \\
$-$0.65 0.5  &     0.76 0.91 1.0 &     0.99 1.19 1.0 &     1.75 2.10 1.0 &     1.41 1.76 0.2 &     1.53 1.94 0.2 &     0.70 0.83 1.0 &     2.41 2.84 0.4 &     2.53 3.04 0.4 &     1.54 2.14 1.0 &     3.18 3.75 1.0 &   3.30 3.95 1.0 \\
$-$0.15 0.5  &     0.83 0.97 1.0 &     1.03 1.24 1.0 &     1.86 2.20 1.0 &     1.54 1.91 1.0 &     1.69 2.13 1.0 &     0.77 0.89 1.0 &     2.68 3.12 1.0 &     2.82 3.33 1.0 &     1.75 2.34 1.0 &     3.52 4.06 1.0 &   3.66 4.28 1.0 \\
\hline
$-$1.65 0.75 &     0.65 0.85 0.4 &     0.91 1.16 0.2 &     1.57 2.02 0.2 &     1.14 1.48 0.0 &     1.22 1.70 0.0 &     0.57 0.77 0.3 &     2.04 2.63 0.2 &     2.12 2.80 0.2 &     1.17 1.90 0.2 &     2.69 3.48 0.3 &   2.77 3.64 0.3 \\
$-$1.15 0.75 &     0.70 0.91 1.0 &     0.94 1.19 1.0 &     1.64 2.11 1.0 &     1.27 1.79 0.0 &     1.37 1.97 0.2 &     0.63 0.82 1.0 &     2.19 2.84 0.4 &     2.29 3.05 0.4 &     1.32 2.14 0.2 &     2.89 3.74 0.4 &   2.99 3.95 0.4 \\
$-$0.65 0.75 &     0.75 0.96 1.0 &     0.97 1.23 1.0 &     1.71 2.19 1.0 &     1.39 1.93 0.4 &     1.51 2.15 0.4 &     0.68 0.87 1.0 &     2.39 3.09 0.6 &     2.50 3.31 0.6 &     1.49 2.34 1.0 &     3.14 4.03 1.0 &   3.25 4.26 1.0 \\
$-$0.15 0.75 &     0.80 1.01 1.0 &     1.01 1.27 1.0 &     1.81 2.28 1.0 &     1.53 2.08 0.6 &     1.67 2.33 1.0 &     0.75 0.93 0.9 &     2.63 3.33 1.0 &     2.77 3.58 1.0 &     1.68 2.51 1.0 &     3.45 4.32 1.0 &   3.59 4.57 1.0 \\
\hline
 & \multicolumn{11}{c}{GC simulations including 10\% color scatter} \\
\tableline
$-$1.65 0.25 &     0.66 0.66 0.0 &     0.91 0.91 0.0 &     1.56 1.58 0.0 &     1.13 1.17 0.0 &     1.22 1.25 0.0 &     0.59 0.60 0.0 &     2.05 2.07 0.0 &     2.13 2.29 0.2 &     1.16 1.20 0.0 &     2.70 2.74 0.0 &   2.79 3.01 0.0 \\
$-$1.15 0.25 &     0.72 0.73 0.0 &     0.95 0.96 0.0 &     1.66 1.69 0.0 &     1.24 1.35 0.9 &     1.34 1.47 1.0 &     0.66 0.67 0.0 &     2.20 2.36 1.0 &     2.29 2.48 1.0 &     1.31 1.50 0.6 &     2.89 3.12 1.0 &   2.99 3.23 1.0 \\
$-$0.65 0.25 &     0.81 0.82 0.0 &     1.02 1.14 1.0 &     1.80 1.99 1.0 &     1.42 1.44 0.0 &     1.55 1.56 0.0 &     0.74 0.77 0.1 &     2.43 2.63 1.0 &     2.56 2.80 1.0 &     1.58 1.85 1.0 &     3.21 3.48 1.0 &   3.34 3.64 1.0 \\
$-$0.15 0.25 &     0.87 0.96 1.0 &     1.08 1.21 1.0 &     1.97 2.16 0.4 &     1.56 1.75 0.4 &     1.72 1.94 1.0 &     0.84 0.86 0.0 &     2.74 2.94 1.0 &     2.89 3.14 0.9 &     1.87 2.21 1.0 &     3.60 3.87 0.9 &   3.76 4.07 1.0 \\
\hline
$-$1.65 0.5  &     0.66 0.78 0.2 &     0.91 1.06 0.0 &     1.57 1.81 0.2 &     1.11 1.26 0.9 &     1.21 1.45 0.2 &     0.57 0.68 0.4 &     2.04 2.37 0.2 &     2.12 2.49 0.3 &     1.17 1.54 0.2 &     2.70 3.15 0.3 &   2.78 3.27 0.3 \\
$-$1.15 0.5  &     0.70 0.85 0.4 &     0.95 1.14 0.4 &     1.66 2.00 0.4 &     1.25 1.43 0.9 &     1.38 1.70 0.0 &     0.63 0.77 1.0 &     2.20 2.59 0.4 &     2.30 2.76 0.4 &     1.33 1.85 0.4 &     2.90 3.44 0.4 &   3.00 3.60 0.4 \\
$-$0.65 0.5  &     0.76 0.91 1.0 &     0.99 1.19 1.0 &     1.75 2.09 1.0 &     1.41 1.78 0.2 &     1.53 1.96 0.2 &     0.70 0.84 1.0 &     2.41 2.84 0.6 &     2.53 3.05 0.4 &     1.54 2.14 1.0 &     3.18 3.74 1.0 &   3.30 3.96 1.0 \\
$-$0.15 0.5  &     0.83 0.98 1.0 &     1.03 1.23 1.0 &     1.86 2.21 1.0 &     1.54 1.91 0.6 &     1.69 2.14 1.0 &     0.78 0.90 0.8 &     2.68 3.11 1.0 &     2.82 3.34 1.0 &     1.75 2.34 1.0 &     3.51 4.05 1.0 &   3.67 4.29 1.0 \\
\hline
$-$1.65 0.75 &     0.66 0.86 0.3 &     0.91 1.15 0.3 &     1.57 2.01 0.3 &     1.15 1.56 0.0 &     1.23 1.76 0.0 &     0.57 0.76 0.4 &     2.04 2.63 0.2 &     2.12 2.80 0.2 &     1.18 1.91 0.2 &     2.69 3.48 0.3 &   2.78 3.65 0.3 \\
$-$1.15 0.75 &     0.70 0.92 0.6 &     0.94 1.20 0.6 &     1.64 2.11 1.0 &     1.27 1.82 0.0 &     1.37 2.00 0.2 &     0.62 0.82 1.0 &     2.19 2.84 0.4 &     2.29 3.05 0.4 &     1.32 2.14 0.4 &     2.89 3.75 0.4 &   3.00 3.95 0.4 \\
$-$0.65 0.75 &     0.75 0.96 1.0 &     0.97 1.23 1.0 &     1.72 2.19 1.0 &     1.40 1.94 0.3 &     1.51 2.15 0.4 &     0.68 0.88 1.0 &     2.38 3.09 0.6 &     2.50 3.31 0.6 &     1.49 2.34 1.0 &     3.14 4.03 1.0 &   3.26 4.27 1.0 \\
$-$0.15 0.75 &     0.80 1.01 1.0 &     1.01 1.27 1.0 &     1.80 2.28 1.0 &     1.52 2.08 1.0 &     1.67 2.33 1.0 &     0.76 0.94 0.9 &     2.62 3.33 1.0 &     2.77 3.59 1.0 &     1.68 2.51 1.0 &     3.45 4.31 1.0 &   3.60 4.57 1.0 \\
\enddata
\tablenotetext{a}{For each color, the location of the blue and red color peaks is reported, together with the value  $P_b$ which represents the likelihood that the color distribution is better fitted by a bimodal color distribution [in the text referred to as $P(bimodal)$].}
\label{tab_spot}
\end{deluxetable}